# An Exploration into Web Session Security- A Systematic Literature Review

MD. IMTIAZ HABIB, 19-39389-1@student.aiub.edu, American International University- Bangladesh

ABDULLAH AL MARUF, 19-39687-1@student.aiub.edu, American International University- Bangladesh

MD. JOBAIR AHMAD NABIL, 18-38837-3@student.aiub.edu, American International University- Bangladesh

**Abstract:** The most common attacks against web sessions are reviewed in this paper, for example, some attacks against web browsers' honest users attempting to create session with trusted web browser application legally. We have assessed with four different ways to judge the viability of a certain solution by reviewing existing security solutions which prevent or halt the different attacks. Then we have pointed out some guidelines that have been taken into account by the designers of the proposals we reviewed. The guidelines we have identified will be helpful for the creative solutions proceeding web security in a more structured and holistic way.

## 1 INTRODUCTION

The web is the most common way to get data and application. It is incredibly complicated and diverse, since it incorporates a plethora of dynamic material created by many parties in order to provide the best possible user experience. Because of putting in place novel security mechanisms typically prevents existing websites from working correctly or negatively affects the user experience, which is generally regarded as unacceptable, given the massive user base of the Web, this heterogeneity makes it very difficult to effectively enforce security. However, the constant pursuit of usability and backward compatibility has had a subtle impact on web security research: new defensive mechanism designers have been extremely careful, and the vast majority of their suggestions consist of relatively local remedies against very particular attacks. As seen by the growth of numerous threat models against which different ideas have been tested, often with rather distinct underlying assumptions, this piecemeal evolution hampered a comprehensive knowledge of many nuanced vulnerabilities and difficulties. It's easy to become lost amid the many offered solutions, and it's nearly hard to appreciate the relative merits and downsides of each one without a comprehensive understanding of the current literature. In this work, we tackle the difficult challenge of providing a comprehensive review of a broad class of frequent attacks that target the contemporary Web, as well as the security solutions that have been presented thus far. We did this research concerning on web session attacks, which are attacking honest web browser users who are attempting to create an authenticated connection with a trustworthy online application. This type of attack takes use of the Web's inherent complexity by interfering with dynamic content, client-side storage, or cross-domain connections, for example, in order to alter browser activity and communication of network. Our decision is based on the fact that web session attacks represent a significant subgroup of important online security events and a variety of solutions,. Then we look at existing security solutions and processes for preventing or mitigating various threats, and we assess each proposal in terms of the security guarantees it offers. When security can only be assured if certain assumptions are met, we state them explicitly. We also consider the influence of each security solution on compatibility and usability, as well as implementation simplicity. These are crucial factors for determining the feasibility of a solution, as well as for determining to what extent each solution, in its current condition, is suitable to wide-scale adoption over the Internet. We also offer an overview of them in a separate section because there are various ideas in the literature that try to provide adequate defenses against various attacks. We detail the assaults each of these ideas prevents in relation to the attacker model considered in its original design, and we evaluate their appropriateness using the criteria outlined above. for computer security researchers we propose an overview of security-relevant components of the web browser and the security policies based on these components, we also show how well-known enforcement techniques are applied in a web browser setting [1]. Lastly, we compile a list of five criteria based on our findings, each of which has been considered to varying degrees by the designers of the various solutions. We note that none of the existing plans adhere to all of the standards, and we suggest that this is due to the Web's inherent complexity and difficulties in protecting it. Security-Critical services are more and more supplied online today and this increases the need of effective defenses for the web platform [3]. We hope that these principles will aid in the creation of novel online security solutions that are more methodical and thorough.

## 1.1 Scope of the survey

1.1 Scope of the survey: Web security is complicated, as web sessions can be targeted at several levels. It is thus necessary to examine some of the assumptions we make and their implications for security in order to clarify the scope of the current study.

• Web sessions can be affected at the network layer by network sniffing or man-in-the-middle attacks. The HTTPS protocol encrypts web traffic and wraps it in an SSL/TLS encrypted channel to safeguard it. We do not consider cryptographic protocol attacks. We presume, in particular, that the attacker will be unable to decrypt, change, or inject data delivered across an encrypted channel to a trustworthy web application. However, we do not assume that web developers always configure HTTPS correctly, as this is a delicate issue that needs to be highlighted in this survey.

• The attacker has not infiltrated the web browser: Web applications frequently rely on conventional web browser protection measures such as the same-origin policy and the HttpOnly cookie attribute. We assume that all of these measures work as expected and that the attacker does not use browser vulnerabilities; otherwise, even secure web apps would be vulnerable. To counter these attacks, modern web browsers implement native cookie protection mechanisms based on the HttpOnly and Secure flags [4].

• Content injection vulnerabilities may harm trustworthy online applications: this is a cautious assumption, since experience has shown that it is very hard to ensure that a web application is free of such dangers. We concentrate on content injection flaws that affect the web browser, such as cross-site scripting assaults (XSS). Content injections that affect the web application's back-end, such as SQL injections, are not supported.

## 1.2 Structure of the Survey

Section 2 provides the background of the study and main structure of the Web. Section 3 is all about methodology of the research, how this research is conducted and what are the questions of this research. Section 4 portrays discussion and the research questions. Section 5 describes the future research possibility from this research. Section 6 explains the validity threat of the research. Section 7 is the conclusion of the research and Section 8 is about the contributions on making this paper.

## 2 RESEARCH BACKGROUND

Here is a quick outline of the fundamental structure of the web and its respective security foundations.

### 2.1 Languages

The data of web pages are stored on server and are recognized by Uniform Resource Location (URL). URL gives the privilege of excessing both location of a resource and criteria to retrieve them. Usually Protocol, Host and Path are included in URL. There are languages for structuring web pages like Hyper text Markup Language(HTML) or comparable language and to style the page there is Cascading Style Sheet(CSS). There are also programming languages like JavaScript that allows to develop highly interactive web applications. Programs written in JavaScript can be included internally in the web page or externally directly. User can update pages usually by asynchronous communications with the help of Ajax requests (via the XMLHttpRequest API).

### 2.2 Locating

All the information regarding web pages are hosted on web servers and can be recognized by a Uniform Resource Locator (URL). URL gives the privilege of excessing both location of a resource and criteria to retrieve them. A specific organization controls the Hosting system which is belong to Domain. For instance, www.google.com belongs to the google.com domain.

### 2.3 Hyper Text Transfer Protocol

The Hypertext Transfer Protocol (HTTP) is an application layer protocol in the Internet protocol suite model for distributed, collaborative, hypermedia information systems. HTTP is the foundation of data communication for the World Wide Web, where hypertext documents include hyperlinks to other resources that the user can easily access, for example by a mouse click or by tapping the screen in a web browser. Development of HTTP was initiated by Tim Berners-Lee at CERN in 1989 and summarized in a simple document describing the behavior of a client and a server using the first HTTP protocol version that was named 0.9.That first version of HTTP protocol soon evolved into a more elaborated version that was the first draft toward a far future version 1.0. Development of early HTTP Requests for Comments (RFCs) started a few years later and it was a coordinated effort by the Internet Engineering Task Force (IETF) and the World Wide Web Consortium (W3C), with work later moving to the IETF. HTTP/1 was finalized and fully documented (as version 1.0) in 1996. It evolved (as version 1.1) in 1997 and then its specifications were updated in 1999 and in 2014. Its secure variant named HTTPS is used by more than 76 percent of the websites. HTTP/2 is a more efficient expression of HTTP's semantics "on the wire", and was published in 2015; it is used by more than 45 percent of websites; it is now supported by almost all web browsers (96 percent of users)[8] and major web servers over Transport Layer Security (TLS) using an Application-Layer Protocol Negotiation (ALPN) extension[9] where TLS 1.2 or newer is required. HTTP/3 is the proposed successor to HTTP/2;[12][13] it is used by more than 20 percent of websites;[14] it is now supported by many web browsers (73 percent of users). HTTP/3 uses QUIC instead of TCP for the underlying transport protocol. Like HTTP/2, it does not obsolete previous major versions of the protocol. Support for HTTP/3 was added to Cloudflare and Google Chrome first, and is also enabled in Firefox.

## 2.4 Security

HTTP is a stateless protocol, enabling the communication between a client (front-end) and a server (back-end). Sessions or tokens are used to overcome the stateless nature of HTTP requests. Somehow if cyber attackers gain control over cookies, they can impersonate the user, thereby retrieving their sensitive data. Over the years, web application security began with sessions and now it is based on tokens to improve overall session security. Session security plays a key factor in building secure web applications. A web application is not secure unless it is protected from external attacks like XSS. These malicious scripts are designed to gain access to sensitive data in web applications, including cookies, as they act as a key to store session tokens. Attackers can exploit and gain unauthorized access to the web application because of the improper implementation of authorization or authentication. According to OWASP (Open Web Application Security Project) Top 10, broken authentication is the second biggest risk to web application security.

# 3   RESEARCH METHODOLOGY

In this work, we tackle the difficult challenge of providing a comprehensive review of a broad class of typical assaults targeting the contemporary Web, as well as the security solutions that have been presented thus far. We concentrate on web session attacks, that is, attacks that target honest web browser users who are attempting to create an authenticated connection with a trustworthy online application. An online attacker can inject harmful content into trusted web apps using this broad class of assaults. Content injections can be carried out in a variety of methods, but they are always facilitated by a lack of or incorrect sanitization of some attacker-controlled input in the online application, whether on the client or server-side. These attacks are commonly associated with Cross-Site Scripting (XSS), which involves the injection of malicious JavaScript code; nevertheless, a lack of sufficient filtration might harm HTML content or even CSS rules.

## 3.1 Research Objective

The present approach for Web session management is exchanging cookie/s between a Web application's client and server. Since the inception of the cookie mechanism, the problem of user privacy has been a source of contention. The use of cookies and HTTP headers such as Referrer allows a third party to create user profiles and ascertain, for a large group of users, which sites they visit, when they visit them, and so on. Several warnings regarding a major loss of privacy have been published since the proposal for and establishment of cookies, including one from one of the writers of the RFC, which specified cookies as part of the HTTP headers for session management [5]. This variability makes it difficult to properly enforce security since implementing new security mechanisms sometimes breaks current websites or has a detrimental impact on the user experience, which is widely seen as unacceptable given the vast user base of the Internet. This constant pursuit for usability and backward compatibility, however, has a subtle influence on online security research: new defensive mechanism designers have been quite careful, and the vast majority of their suggestions consist of relatively local remedies against very specific threats. As seen by the growth of numerous threat models against which different ideas have been assessed, often with quite different underlying assumptions, this piecemeal evolution hampered a comprehensive knowledge of many nuanced vulnerabilities and difficulties [6].

## 3.2 Research Questions

RQ1: where I can get a dataset for predicting security flaws?

RQ2: What are the many types and levels of online security? What basic web security elements should be provided for an undergraduate project?

RQ3: What are the advantages and disadvantages of using this strategy for security rather than dependability?

## 3.3 Article Selection

A search string is a combination of keywords, truncation symbols, and boolean operators that you type into the search box of a library database or search engine. A total of 12 significant research lists were generated during the initial phase of the search procedure. The whole text of the 12 significant studies was then scrutinized. The relevance of the data to the study topic and the quality of the research were taken into account during the analysis [7].

*3.3.1 Keywords and Search String.* The words you type into the database search fields are known as keywords, sometimes known as search phrases. We search Research Gate, Science Direct, ACM, and Google Scholar using those keywords. When we search with our broadly automotive keyword then we find an initial set of articles. In every article, we read the full article title and abstract carefully. For data collecting, digital libraries are open-source solutions. Find a relevant paper for this paper using those keywords [8].

*3.3.2 Digital Libraries to Search.* Searching digital libraries entail looking for information in remote databases of digitized or digital materials and obtaining it. These databases might contain the metadata for a particular object of interest. In digital libraries, we search by title and time duration. From the main library page, we are going to the Advanced Search link. Begin by typing in a single search word, such as social media. Then, if you place quote marks around terms, you may gradually concentrate in on the most relevant results.

*3.3.3 keyword search and Manual Selection.* Keywords are the words and phrases that users type into search engines to find what they're looking for." Web Session Security", "The Formal Foundation of the Web", "Micro-Policies Security", and "Improving Session Security" are some of the search keywords we used in our article. We are completing this research from 9 several journals and articles [8].

*3.3.4 Final set of Articles.* At the ending, we select 9 articles for our topic from the 12 articles which are suitable for our topic. Our selected topics are very much familiar with our topics we are following those articles for completing our research paper. We refer to those articles in our reference section.

## 4  DISCUSSION

A number of well-designed suggestions are now apparent that meet many of the criteria we established. All of the proposals fail to fulfill all of the requirements. Web security is a challenging problem, not because of the standards' nature, but because the world's largest distributed system requires a wide range of considerations at many different levels. In the full version of the poll, we explain how the particular difficulties of the web platform have affected the evolution of online security research. The adoption of formal techniques from the outset of the design process is also something we recommend for novel online security solutions using hybrid client/server architectures. Formal techniques may be used to enhance internet security, according to a recent study [9]. The evaluation offers potential research avenues. See [10] for an overview of outstanding difficulties and possible research areas in the quest for trustworthy online session security solutions.

## 4.1 RQ1

Transparency. Our definition of transparency is a mix of good usability and total compatibility: we feel that, given the Web's massive user base and heterogeneity, this is the most crucial component for guaranteeing a large-scale deployment of any defense system. Because usability issues may deactivate or otherwise bypass current security measures, it is widely accepted that security comes at the price of usability [13]. Without the need to update browsers, both web developers and users can benefit from cutting-edge security mechanisms without noticeable changes in the content of their websites. This is especially important for users who do not want to update their browsers and for web developers who don't want to adopt new defensive technologies.

## 4.2 RQ2

Verification and specification in the strictest sense. Formal models and approaches have lately been used in the development and testing of new ideas for online session security. While a formal specification is of little benefit to web developers, it helps security specialists understand the complexities of the proposed solution more clearly. A rigorous definition of semantic security, such as non-interference [10] or session integrity, may inspire web security designers instead of giving ad-hoc solutions to the multitude of low-level attacks that presently target the Web.

## 4.3 RQ3

The precision of the algorithm may aid investors in avoiding losses while simultaneously increasing their earnings. The researchers employed a variety of methodologies, including linear regression, neural networks, evolutionary algorithms, SVM, K-Nearest Neighbor, and Random Forest, to achieve their results.

## 5  FUTURE RESEARCH DIRECTIONS

Based on the results of this systematic literature review, we propose that in the future, we employ a mix of input data set types for stock prediction in order to improve accuracy. Combining multiple different approaches for stock market prediction may help to increase the accuracy of machine learning. Machine learning algorithms that make use of ensemble methods include boosting, deep learning, and ensemble approaches, to name a few examples. The three elements that we believe are responsible for the various study gaps in stock prediction performance are: method selection, data set type selection, and machine learning performance modification strategies.

## 5.1 Topic-1: Machine Learning

In order to get superior results, it is possible to combine many machine learning algorithms. Combining SVM with LSM, for example, may improve the accuracy of a system by increasing its precision [10].

## 5.2 Topic-2: Models

An ensemble approach to generating a strong classifier from weaker ones is known as "boost." A training dataset is used to build a model, and then a second model is built to correct the previous model's errors.

### 5.3 Topic-3: Deep Learning

Deep learning algorithms such as ANN, RNN, and other neural network algorithms may be used to predict stock market values by analyzing data from a variety of sources such as websites, social media platforms, and financial news websites. These models are quite useful in today's society, since the market's value varies in response to news reported on social media [11].

## 6 VALIDITY THREAT

The Prediction System may encounter a slew of obstacles and roadblocks along the way. It is possible that biased news on an online website will cause a significant shift in the prediction system, necessitating the development of a fake news detection algorithm. Algorithms are unable to precisely forecast the movement of the stock market. When doing research, it is important to keep these issues in mind for the researcher.

## 7 CONCLUSION

Throughout the years, there have been several online session hacks discovered. In addition, we looked at some of the most often used defensive strategies. For We examined the security guarantees provided by each solution in the face of a wide range of threats. There is a low barrier to entry for attacker models, their influence on usability and compatibility, and the repercussions of these models. In the last several years, researchers have offered formal approaches as a helpful tool for building and evaluating online security solutions, which we have thoroughly reviewed. Formal techniques have been successfully applied to a variety of online security concerns, including JavaScript, browser, web application, and web protocol analysis, to name a few examples. Following our investigation of security conferences over the preceding few years, we have come to the conclusion that no new protocols have been disclosed that have not been carefully reviewed for security risks.

## REFERENCES

[1] Nataliia Bielova. 2013. Survey on JavaScript Security Policies and their Enforcement Mechanisms in a Web Browser. Journal of Logic and Algebraic Programming 82, 8 (2013), 243–262.

[2] Aaron Bohannon and Benjamin C. Pierce. 2010. Featherweight Firefox: Formalizing the Core of a Web Browser. In USENIX Conference on Web Application Development, WebApps 2010.

[3] Michele Bugliesi, Stefano Calzavara, and Riccardo Focardi. 2017. Formal methods for web security. Journal of Logical and Algebraic Methods in Programming (2017). To appear.

[4] Michele Bugliesi, Stefano Calzavara, Riccardo Focardi, and Wilayat Khan. 2015. CookiExt: Patching the Browser Against Session Hijacking Attacks. Journal of Computer Security 23, 4 (2015), 509–537.

[5] Towards a formal Foundation of web security. (n.d.). Retrieved from https://brambonne.com/docs/Thesis.pdf

[6] Web security. (n.d.). Retrieved from https://www.researchgate.net/topic/Web-Security

[7] Micro-policies for web session security. (n.d.). Retrieved from https://ieeexplore.ieee.org/abstract/document/7536375

[8] Towards a formal Foundation of web security. (n.d.). Retrieved from https://ieeexplore.ieee.org/abstract/document/5552637

[9] Michele Bugliesi, Stefano Calzavara, and Riccardo Focardi. 2017. Formal methodsfor web security. Journal of Logical and Algebraic Methods in Programming 87(2017), 110–126.

[10] Mary Frances Theofanos and Shari Lawrence Pfleeger. 2011. Guest Editors' Introduction: Shouldn't All Security Be Usable? IEEE Security Privacy 9, 2 (2011), 12–17.

[11] Willem De Groef, Dominique Devriese, Nick Nikiforakis, and Frank Piessens. 2012.Flow Fox: a Web Browser with Flexible and Precise Information

[12] Flow Control. In Proceedings of the 19th ACM Conference on Computer and Communications Security, CCS 2012. 748–759

[13] Jingyi Shen and M. Omair Shafiq. 2020. Short-term stock market price trend prediction using a comprehensive deep learning system. Journal of Big Data 42, 66 (Aug. 2020). https://journalofbigdata.springeropen.com/articles/10.1186/s40537-020-00333-6

[14] TarjaSysta, M.M. MahbubulSyeed, and ImedHammouda. 2013. Evolution of Open Source Software Projects: A Systematic Literature Review. Journal of Software 8, 11 (Nov. 2013). https://doi.org/10.4304/jsw.8.11.2815-2829.

## 8   CONTRIBUTION RECORD

Here are the contribution information on making this research paper.

### 8.1 Paper Assessment

The following table includes the contributions of individual group member on the paper assessment

| Student Id and Name | Paper No from Ref | Paper Title |
|---|---|---|
| MD. IMTIAZ HABIB (19-39389-1) | 1,2,3,4 | Survey on JavaScript Security Policies and the Enforcement Mechanisms in a Web, Feather weig Firefox: Formalizing the Core of a Web Browse Formal methods for web security, CookiExt: Patchir the Browser Against Session Hijacking Attacks. |
| ABDULLAH AL MARUF (19-39687-1) | 5,6,7,8 | Towards a formal Foundation of web security, We security, Micro-policies for web session securit Towards a formal Foundation of web security. |
| MD. JOBAIR AHMAD NABIL (18-38837-3) | 10,11,12,14,15 | Formal methods for web security, Guest Editor Introduction: Shouldn't All Security Be Usable? Flo Fox: a Web Browser with Flexible and Preci Information Flow Control, Short-term stock mark price trend prediction using a comprehensive dee learning system, Evolution of Open Source Softwa Projects: A Systematic Literature Review |

Table 1. Paper collected and read by the group member

## 8.2 Paper writing contribution

The following table includes the contributions of individual group member on writing the paper.

| Student Id and Name | Section No | Section Title |
|---|---|---|
| MD. IMTIAZ HABIB (19-39389-1) | 1,2,3,4 | Research Methodology, Contribution Record |
| ABDULLAH AL MARUF (19-39687-1) | 5,6,7,8 | Research Methodology |
| MD. JOBAIR AHMAD NABIL (18-38837-3) | 10,11,12,14,15 | Discussion, Future Research Directions, Validity Threat |

Table 2. Section(s) Written in the paper by the group member